\documentclass[sigplan,screen]{acmart}
\usepackage{array}
\usepackage{listings}
\usepackage{tikz}
\usetikzlibrary{arrows.meta,positioning}
\usepackage[capitalise,noabbrev]{cleveref}
\definecolor{dtCodeBg}{HTML}{F7F7F7}
\definecolor{dtComment}{HTML}{555555}
\definecolor{dtString}{HTML}{005A8D}
\definecolor{dtKeyword}{HTML}{8A3A00}
\definecolor{dtEval}{HTML}{E8F1FA}
\definecolor{dtTrain}{HTML}{FFF1D8}
\definecolor{dtStudent}{HTML}{EAF7EA}
\definecolor{dtTeacher}{HTML}{F2ECFA}
\definecolor{dtArtifact}{HTML}{F7F7F7}

\lstdefinelanguage{debugjson}{
  morecomment=[l]{\#},
  morestring=[b]",
  stringstyle=\color{dtString},
  commentstyle=\color{dtComment},
  emph={type,source,mode,summary,payload,checkpointKind,text},
  emphstyle=\color{dtKeyword}\bfseries
}

\lstdefinestyle{artifactlisting}{
  basicstyle=\ttfamily\fontsize{6.5}{7.0}\selectfont,
  columns=fullflexible,
  breaklines=true,
  breakatwhitespace=false,
  backgroundcolor=\color{dtCodeBg},
  frame=single,
  framerule=0.2pt,
  rulecolor=\color{dtComment},
  xleftmargin=1pt,
  xrightmargin=1pt,
  aboveskip=2pt,
  belowskip=2pt
}

\newcommand{\myparagraph}[1]{\vspace{2pt}\noindent\textbf{#1}}

\AtBeginDocument{%
}

\setcopyright{cc}
\setcctype{by-nc-nd}
\acmDOI{10.1145/3837729.3840500}
\acmYear{2026}
\copyrightyear{2026}
\acmISBN{979-8-4007-2910-2/2026/10}
\acmConference[SPLASH Companion '26]{Companion Proceedings of the 2026 ACM SIGPLAN International Conference on Systems, Programming, Languages, and Applications: Software for Humanity}{October 3--9, 2026}{Oakland, CA, USA}
\acmBooktitle{Companion Proceedings of the 2026 ACM SIGPLAN International Conference on Systems, Programming, Languages, and Applications: Software for Humanity (SPLASH Companion '26), October 3--9, 2026, Oakland, CA, USA}
\acmSubmissionID{splashcomp26demo-p43-p}
\received{2026-06-27}
\received[accepted]{2026-07-25}

\begin{document}

\title{DebugTracker: Lightweight Process Evidence for Classroom Debugging}

\author{Jiatong Liu}
\orcid{0009-0000-5040-8272}
\affiliation{%
  \institution{Monash University}
  \city{Melbourne}
  \country{Australia}}
\email{jliu0455@student.monash.edu}

\author{Xue Yao}
\orcid{0009-0001-5818-5912}
\affiliation{%
  \institution{Monash University}
  \city{Melbourne}
  \country{Australia}}
\email{xyao0028@student.monash.edu}

\author{Zehua Zhang}
\orcid{0009-0003-2132-0361}
\affiliation{%
  \institution{Monash University}
  \city{Melbourne}
  \country{Australia}}
\email{zzha0610@student.monash.edu}

\author{Yongqiang Tian}
\correspondingauthor
\orcid{0000-0003-1644-2965}
\affiliation{%
  \institution{Monash University}
  \city{Melbourne}
  \country{Australia}}
\email{yongqiang.tian@monash.edu}

\begin{abstract}
Debugging exercises are often assessed from final code and test outcomes, yet
these artifacts hide how students reproduced failures, formed hypotheses,
inspected evidence, edited code, and verified fixes. We present DebugTracker, a
Visual Studio Code extension that records lightweight debugging-process evidence
for classroom tasks. DebugTracker separates uncoached Evaluation Mode traces
from coached Training Mode traces, stores append-only JSONL events, and exports
timeline and Markdown reports for human review. The prototype records test
commands, editor and debugger metadata, student checkpoints, source snapshots,
and human labels. It can also record optional image evidence and AI-assisted
practice feedback.
DebugTracker is largely language-agnostic: it captures process
evidence through standard VS Code mechanisms rather than language-specific
tooling, although debugger evidence depends on the relevant VS Code language
extension. We validate the prototype with debugging tasks in Python, TypeScript,
and Java, 16 automated checks, and an 11-case manual trial matrix spanning
packaged VSIX installation and three operating systems.
\end{abstract}



\maketitle

\section{Introduction}

Debugging is a process of observing failures, asking questions, inspecting
state, testing hypotheses, and validating repairs~\cite{sillito2006questions,
zeller2009why}. It is also one of the hardest skills for students to learn
because the reasoning that separates a systematic investigation from lucky
guessing happens between the failing test and the final patch. Classroom
assessment, however, commonly considers only that final patch, the final test result,
or a short written note. The intermediate steps---reproducing the failure,
narrowing it to a suspect location, forming a hypothesis, and checking that the
repair fixes the observed symptom---leave no durable record once the session
ends.

This creates a process-visibility gap in assessment. For a student, a debugging
exercise unfolds as the rich, multi-step process just described. For an
instructor or examiner, however, the available artifact is often only the
submitted code plus a pass/fail test result. Consider
two students who both submit a passing solution: one may have followed an
evidence-based debugging process, while the other may have edited and tested
repeatedly until the suite turned green. The final code can look identical even
when the underlying debugging competence differs. When the goal of the exercise
is to teach a process, feedback that ignores the process misses what the
exercise was meant to assess.

Richer alternatives exist but are not well suited to the classroom. Screen
recording and live observation capture the process faithfully, yet they do not
scale to large
cohorts, are intrusive, and collect far more data---faces, unrelated windows,
keystrokes---than a review task requires. Commit history can support debugging
tasks such as locating failure-inducing changes through bisection and
deduplicating compiler bugs~\cite{zhou2026bisection}. However, it is too coarse
and is easily rewritten, so it rarely reflects the moment-to-moment
investigation needed for classroom assessment.
Automated debugging and fault-localization tools target the act of finding bugs,
not the assessment of how a learner did so, and prior work cautions that such
automation does not straightforwardly help programmers in
practice~\cite{parnin2011debugging}. Classroom assessment therefore needs a
lightweight, structured record of the debugging \emph{process}: one that is
cheap to produce, cheap to read, and bounded to the task at hand.

DebugTracker targets this gap with a task-scoped IDE tool. Its goal is not to
replace debuggers, automated fault localization, or interactive explanation
systems such as the Whyline~\cite{ko2004whyline}; nor does it claim that
automated debugging alone can assess student practice.
Instead, it makes the evidence surrounding ordinary tests, breakpoints, edits,
checkpoints, and verification inspectable after a session ends, so an instructor
can review the debugging narrative rather than only its outcome.

\begin{figure*}[t]
  \centering
  \resizebox{0.9\linewidth}{!}{%
  \begin{tikzpicture}[
    labelbox/.style={fill=white, fill opacity=0.93, text opacity=1,
      draw=black!20, rounded corners=2pt, inner sep=2.5pt, align=center},
    modelabel/.style={labelbox, font=\sffamily\bfseries\normalsize},
    zonelabel/.style={labelbox, font=\sffamily\bfseries\small},
    steplabel/.style={labelbox, font=\sffamily\footnotesize, inner sep=1.6pt,
      text width=1.9cm},
    artifactlabel/.style={labelbox, font=\sffamily\footnotesize, inner sep=1.6pt}
  ]
    \node[anchor=south west, inner sep=0] (base) at (0,0)
      {\includegraphics[width=\textwidth]{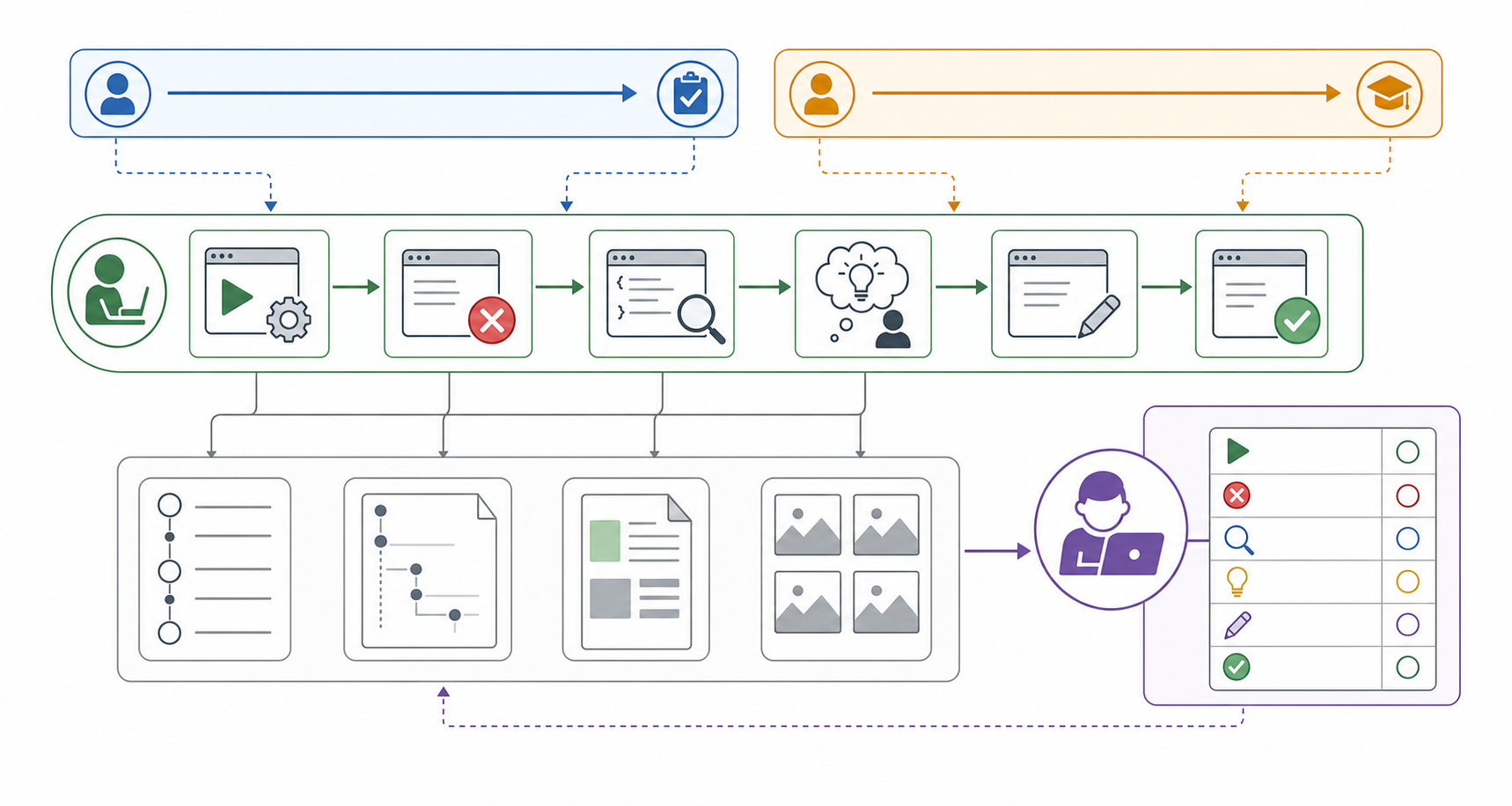}};
    \begin{scope}[x={(base.south east)}, y={(base.north west)}]
      \node[modelabel] at (0.265,0.92) {Evaluation Mode\\
        \normalfont uncoached assessment};
      \node[modelabel] at (0.725,0.92) {Training Mode\\
        \normalfont coached practice};
      \node[zonelabel] at (0.145,0.705) {Student debugging};
      \node[zonelabel] at (0.205,0.465) {Evidence package};
      \node[zonelabel] at (0.775,0.475) {Instructor review};
      \node[steplabel] at (0.178,0.545) {Run tests};
      \node[steplabel] at (0.335,0.545) {Observe failure};
      \node[steplabel] at (0.465,0.545) {Inspect code};
      \node[steplabel] at (0.595,0.545) {Hypothesize};
      \node[steplabel] at (0.725,0.545) {Edit fix};
      \node[steplabel] at (0.855,0.545) {Verify};
      \node[artifactlabel] at (0.172,0.210) {Event log};
      \node[artifactlabel] at (0.317,0.210) {Timeline};
      \node[artifactlabel] at (0.472,0.210) {Report};
      \node[artifactlabel] at (0.602,0.210) {Snapshots/\\images};
      \node[artifactlabel, font=\sffamily\footnotesize, text width=2.4cm]
        at (0.885,0.130) {Checklist labels};
    \end{scope}
  \end{tikzpicture}%
  }
  \caption{DebugTracker workflow from session mode to reviewable evidence. A
  session begins in one of two mutually exclusive modes (top). \emph{Evaluation
  Mode} (left, blue) records an uncoached assessment trace: hints, solution
  prompts, and after-session feedback are disabled. \emph{Training Mode} (right,
  orange) records coached practice: process prompts and optional AI feedback are
  enabled, but direct patches are still disallowed. Both modes drive the
  \emph{same} debugging path (middle): run tests, observe a failure, inspect
  code, state a hypothesis, edit the fix, and verify the result. DebugTracker
  captures this path as an event log, timeline, report, snapshots or images, and
  human labels (bottom), giving instructors a process-oriented basis for feedback
  while keeping assessment traces and coached practice on separate, clearly
  labeled tracks.}
  \Description{Illustrated workflow showing Evaluation Mode and Training Mode
  feeding into a student debugging path, which produces evidence artifacts for
  instructor review.}
  \label{fig:workflow}
\end{figure*}

DebugTracker is designed for three users: students who need a lightweight way to
record their debugging narrative, instructors and teaching assistants who need
reviewable evidence, and researchers who need structured traces without full
screen capture. The paper contributes (1) a VS Code extension implementing this
workflow, (2) a privacy-conscious event and report model for debugging-process
evidence, and (3) a validation package with cross-language sample tasks,
automated tests, a demo script, and a packaged VSIX build. A demonstration video
is available at \url{https://youtu.be/-hzR-Whu2s4}.

\section{DebugTracker}

\cref{fig:workflow} summarizes the workflow that DebugTracker makes
visible. A session begins by selecting a mode: Evaluation Mode produces an
uncoached assessment trace, whereas Training Mode records coached practice.
In either mode, the student carries out the same ordinary debugging routine
inside VS Code---the path shown across the middle of \cref{fig:workflow},
from running tests to verifying the fix. \Cref{sec:demo} presents a concrete
example. From the instructor's perspective, those same actions
become a structured evidence package: an event log, timeline, report, snapshots
or images, and optional human labels.

The command surface is deliberately small: start an Evaluation session, start a
Training session, add a checkpoint---a timestamped failure observation,
hypothesis, or verification note---and finish. The two modes differ only in
coaching. Evaluation Mode is uncoached: it disables process prompts, solution
hints, and after-session feedback, and marks its reports as assessment traces.
Training Mode is coached: it provides prompts during debugging---for example,
reminding the student to reproduce the failure, record a
hypothesis before editing, and verify the fix afterward---and can add optional,
OpenAI-compatible feedback on \emph{how} the student is debugging. This coaching
deliberately targets the process, not the answer: it never reveals the fix or
patches the code for the student. The explicit mode boundary keeps these coached
practice traces separate from assessment evidence.

When the session ends, DebugTracker exports a Markdown report and a timeline that
together summarize the session: its mode and privacy boundary, the detected tests
and failures, the student's hypotheses and edits, and the post-edit verification.
Instructors can add human labels afterward, and those labels are stored as
separate events rather than overwriting the raw trace.

\section{Demonstration}
\label{sec:demo}

This section describes the live demonstration: a single scripted debugging
session that an audience can follow end to end, and the assessment evidence it
leaves behind.

The demonstration uses the included Python checkout-pricing task;
TypeScript and Java variants are also available. The task includes a failing
test caused by a bug that is small enough to fix on stage but still demands a
genuine debugging chain: the
shipping-eligibility check uses the original subtotal instead of the discounted
taxable subtotal, so a discounted order is incorrectly granted free shipping.

\cref{tab:demo} lists the script and, for each step, the assessment question
that the step's recorded evidence lets an instructor answer. The presenter
installs the VSIX (or launches the extension from source), opens the sample task,
selects the Evaluation Mode path in \cref{fig:workflow}, and runs
\texttt{python -m unittest}. Once the failing test appears, the presenter follows
the student path from the figure: record the observed failure, inspect
\texttt{src/pricing.py}, state the hypothesis that the free-shipping threshold
uses the wrong subtotal, edit the shipping-basis assignment, rerun the test, and
record verification. A final step opens the timeline and Markdown report;
\cref{fig:report} shows an excerpt, including the underlying JSONL events
whose schema is defined in \cref{sec:design}.

The demonstration therefore emphasizes the trace rather than the fix. From the
final patch alone, the two
students contrasted in the introduction are indistinguishable; reading the
exported trace, an instructor can answer every question in the last column of
\cref{tab:demo}---whether the failure was reproduced, whether the suspect code
was inspected, whether the hypothesis preceded the edit, whether the edit was
targeted, whether verification followed it, and whether any coaching was
involved---without having watched the session.

\begin{table}[htbp]
  \caption{Demonstration script and the instructor checks enabled by each step.}
  \Description{Table mapping each demonstration step to the evidence DebugTracker
  captures and the instructor check that evidence enables.}
  \label{tab:demo}
  \small
  \setlength{\tabcolsep}{4pt}
  \renewcommand{\arraystretch}{1.2}
  \begin{tabular}{@{}>{\raggedright\arraybackslash}p{0.22\linewidth}
                    >{\raggedright\arraybackslash}p{0.45\linewidth}
                    >{\raggedright\arraybackslash}p{0.22\linewidth}@{}}
    \toprule
    \textbf{Demo step} & \textbf{Captured evidence} & \textbf{Instructor check} \\
    \midrule
    Start session & Mode, purpose, privacy boundary, and session id. &
    Coaching used? \\
    \addlinespace[2pt]
    Run failing test & Test command and failure context. &
    Failure reproduced? \\
    \addlinespace[2pt]
    Inspect code & Editor navigation and debugger activity. &
    Suspect code inspected? \\
    \addlinespace[2pt]
    Record hypothesis & Pre-edit failure note and hypothesis. &
    Hypothesis before edit? \\
    \addlinespace[2pt]
    Edit fix & Source snapshot and edit metadata. &
    Targeted repair? \\
    \addlinespace[2pt]
    Verify and export & Passing test event, timeline, report, and optional
    labels. & Verification after edit? \\
    \bottomrule
  \end{tabular}
\end{table}

\begin{figure}[t]
\begin{lstlisting}[style=artifactlisting,language=debugjson]
# DebugTracker Student Summary (Evaluation)
Recorded: test yes; checkpoint yes; edit yes; snapshots yes.
Timeline: 02:34 unittest -> 02:35 failure -> hypothesis
  -> 02:36 edit src/pricing.py -> 02:37 verification.

{"type":"failure.observed","source":"student","mode":"evaluation",
 "summary":"discounted order wrongly gets free shipping",
 "payload":{"checkpointKind":"failure"}}
{"type":"student.hypothesis","source":"student","mode":"evaluation",
 "summary":"free-shipping threshold uses original, not discounted, subtotal",
 "payload":{"checkpointKind":"hypothesis"}}
{"type":"verification.recorded","source":"test","mode":"evaluation",
 "summary":"unittest passes after fix","payload":{"checkpointKind":"verification"}}
\end{lstlisting}
  \caption{Exported Markdown report excerpt and syntax-highlighted JSONL event
  excerpts from the Python \texttt{checkout-pricing} evaluation task.}
  \Description{Listing with a report excerpt and three JSONL events for failure,
  hypothesis, and verification.}
  \label{fig:report}
\end{figure}

The same artifacts support a lightweight classroom workflow beyond the
demonstration. An
instructor publishes a debugging task with its source, tests, and any custom
test-command patterns; students open it in VS Code, run an Evaluation session,
debug normally, and submit the exported report and trace alongside their final
solution. An instructor or teaching assistant then reviews the process evidence directly,
without screen recordings or commit-history archaeology. Because the questions in
\cref{tab:demo} map onto common debugging rubrics, the reviewer can attach
rubric labels through the human-label mechanism. These labels are stored as later
events, leaving the student's original trace intact.

\section{Design and Implementation}
\label{sec:design}
DebugTracker is implemented in TypeScript as a VS Code extension targeting VS
Code 1.88 or newer, using the VS Code extension API~\cite{vscodeapi}. It contributes
a DebugTracker activity-bar container, a session dashboard, a timeline
tree/webview, command-palette actions, and workspace settings for capture and
storage policy. All primary artifacts are workspace-local: JSONL session logs,
Markdown reports, source snapshots, and optional image attachments.

The event schema in \texttt{src/eventTypes.ts} records a session id, timestamp,
mode, source, event type, optional file location, a short summary, and a
structured payload. \cref{tab:event-schema} shows the implemented fields
and event families. The log is append-only, so raw behavior, human labels, and
AI feedback remain distinguishable during review.

\begin{table}[htbp]
  \caption{Implemented DebugTracker event schema.}
  \Description{Table listing event fields, sources, modes, and event families
  implemented by DebugTracker.}
  \label{tab:event-schema}
  \small
  \begin{tabular}{@{}p{0.27\linewidth}p{0.66\linewidth}@{}}
    \toprule
    Schema item & Implemented values or purpose \\
    \midrule
    Required fields & \texttt{id}, \texttt{timestamp}, \texttt{sessionId},
    \texttt{type}, \texttt{source}, \texttt{mode}, \texttt{summary},
    \texttt{payload}. \\
    Optional location & \texttt{filePath}, \texttt{line}, and \texttt{column}
    when an event is tied to source code. \\
    Modes & \texttt{evaluation} and \texttt{training}. \\
    Sources & \texttt{session}, \texttt{editor}, \texttt{debugger},
    \texttt{terminal}, \texttt{test}, \texttt{student}, \texttt{human},
    \texttt{ai}. \\
    Event families & Session start/stop, file open/edit, debugger lifecycle
    (start, stop, step, breakpoint, continue), terminal commands, test runs,
    source snapshots, failure evidence, image evidence, student
    checkpoint/hypothesis/note, verification, human labels, AI analysis, and
    AI coaching. \\
    Payload & Source-specific structured metadata, such as command lines,
    checkpoint text, snapshot paths, image attachment paths, and label details. \\
    \bottomrule
  \end{tabular}
\end{table}

Internally, capture is organized as a thin set of listeners over existing VS Code
events feeding a single append-only writer. When a session starts, the extension
subscribes to editor, debugger, and terminal events, normalizes each one into the
schema of \cref{tab:event-schema}, and appends it to the session's JSONL
log; the timeline tree/webview and the Markdown report are derived views over
that same log rather than independent stores. Test runs are detected from
shell-integrated terminal commands against built-in and instructor-configured
patterns, debugger activity is observed through the Debug Adapter Protocol, and
student checkpoints and human labels are captured through explicit command-palette
actions. Because every artifact is reconstructed from the ordered event stream,
the trace remains the single source of truth: re-rendering a report or timeline
never mutates the underlying evidence, and instructors who review it later see
the same raw sequence the student produced.

DebugTracker is language-agnostic by design. Because it captures evidence
through these standard VS Code mechanisms rather than a language-specific
toolchain, the same event model applies across languages. Built-in patterns
recognize common JavaScript, Python, Java, and C/C++ test commands when shell
integration is available, and instructors can extend coverage through the
\texttt{debugTracker.\allowbreak testCommandPatterns} setting. The languages
used in this paper are therefore a validation set, not a constraint on the tool.

DebugTracker treats privacy as a capture-policy problem: it records
task-relevant metadata rather than a full interaction replay. By default, it
does not store complete terminal output, screen video, mouse movement, or
per-keystroke history. For instructor-authored tasks, it saves baseline and final
source snapshots so instructors can compare the starting bug with the submitted
fix.

Reports also separate captured evidence from instructor-only interpretation.
Student-facing reports describe what was captured and, in Training Mode, provide
practice feedback. Instructor actions come later and never overwrite the trace:
human labels are appended as separate events, and instructor-side reports are
derived from the log. This keeps the student's raw assessment trace distinct from
the instructor's interpretation of it.

\section{Implementation Validation}

This section reports engineering evidence that the implemented workflow installs,
runs, and produces the expected evidence across configurations. It validates the
artifact rather than the pedagogy: it shows that DebugTracker reliably does what
\crefrange{sec:demo}{sec:design} describe, not that the resulting
evidence improves grading. We return to that distinction at the end of the
section.

The automated test entry point \texttt{dist/test/runTests.js} runs 16 checks that
exercise the mechanisms behind those claims: session identifiers, terminal
test-command detection, mode policy, report generation, AI-coach prompt
construction, image evidence, source-snapshot reporting, timeline rendering,
session-summary rendering, human-label entry, and Training Mode feedback. All 16
pass in the current workspace.

A manual validation matrix of 11 documented cases covers cases not exercised
by the unit tests:
packaged VSIX installation; Windows, macOS, and Linux task wrappers; Evaluation
Mode on the Python, TypeScript, and Java tasks; Training Mode feedback; AI-coach
guidance; image evidence; instructor-side report generation; and the
missing-debugger warning. The three language tasks deliberately share a single
intended bug---free-shipping eligibility computed from the original rather than
the discounted taxable subtotal---so that running them confirms DebugTracker
records the debugging process through one shared event model rather than a flow
specialized to a particular language, exercising the language-agnostic design of
\cref{sec:design}.

\myparagraph{Future work.}
The current validation establishes implementation correctness across languages
and platforms, but does not yet demonstrate DebugTracker's usefulness in a real
classroom. We plan a study in which instructors and teaching assistants review
debugging submissions using final-code evidence alone or DebugTracker reports.
We will measure review time, inter-rater agreement, and feedback specificity,
and collect student and instructor perceptions of reporting overhead. We will
also examine whether the recorded process evidence accommodates diverse
debugging strategies rather than encouraging a single prescribed workflow.

\section{Related Work}

DebugTracker complements work in debugging education, programming-process
analytics, and automated debugging.

\myparagraph{Debugging as a learned skill.}
Research identifies debugging as one of the harder competencies for students
to acquire. McCauley et al.~\cite{mccauley2008debugging} survey the educational
literature and treat debugging as a distinct skill rather than a by-product of
writing code, while the multi-institutional study of
Fitzgerald et al.~\cite{fitzgerald2008flailing} reports that novices oscillate
between systematic investigation and ad hoc ``flailing.'' These accounts align
with classic characterizations of debugging as a question-driven,
hypothesis-testing activity~\cite{sillito2006questions, zeller2009why}. Such work
motivates assessing the \emph{process} rather than only the final patch;
DebugTracker operationalizes this by recording the reproduce--observe--hypothesize
--edit--verify chain as reviewable evidence.

\myparagraph{Programming-process data.}
A substantial body of computing-education research captures fine-grained student
activity. Blackbox~\cite{brown2014blackbox} collects snapshots and compilation
events from BlueJ at large scale, Marmoset~\cite{spacco2006marmoset} captures
incremental snapshots through a submission and testing system, and
ProgSnap2~\cite{price2020progsnap2} standardizes a portable format for
programming-process data; surveys of educational data mining catalog how such
traces are analyzed~\cite{ihantola2015edm}. These systems are optimized for
large-scale, often institution-wide, research collection. DebugTracker differs in
granularity and intent: it produces a task-scoped, human-readable narrative for
individual review, records task-relevant metadata rather than full keystroke
replay, and keeps coached and uncoached traces separate.

\myparagraph{Automated debugging and explanation.}
Interactive systems such as the Whyline~\cite{ko2004whyline} let developers ask
why and why-not questions about program behavior, and automated fault
localization has been surveyed extensively~\cite{wong2016fault}. Parnin and
Orso~\cite{parnin2011debugging}, however, caution that such automation does not
straightforwardly help programmers in practice. DebugTracker is complementary to
this line of work: it does not localize faults or explain behavior, but instead
makes the surrounding human investigation inspectable so that an instructor can
judge how a student debugged.

\section{Tool Availability}

DebugTracker is distributed as a self-contained artifact. It contains the VS Code
extension source, a prebuilt VSIX installer, cross-language sample tasks, a demo
script, student and instructor guides, reproducibility notes, and a practice test
report.
The extension targets VS Code 1.88 or newer and installs from the VSIX without
additional services; the optional AI coach is off by default and requires only an
OpenAI-compatible endpoint when explicitly enabled. The source repository is
available at \url{https://github.com/t3-research/DebugTracker}, a
demonstration video is available
at \url{https://youtu.be/-hzR-Whu2s4}, and an archived snapshot of the artifact
is available at \url{https://doi.org/10.5281/zenodo.20955037}. The artifact is
released under CC BY-NC-SA 4.0.

\section{Conclusion}

DebugTracker provides an alternative to both final-code-only grading and
intrusive screen capture. It records enough structured evidence for
instructors to review how a student debugged while keeping coached practice,
human labels, and raw assessment traces separate. By turning ordinary tests,
breakpoints, edits, checkpoints, and verification into a reviewable,
privacy-conscious trace, it makes the debugging process---not just its
outcome---something an instructor can see, discuss, and assess.

\begin{acks}
This research was supported by the Nectar Research Cloud, a
collaborative Australian research platform supported by the NCRIS-funded
Australian Research Data Commons (ARDC). This research was supported by the
Summer and Winter Vacation Research Scholarship Program at Monash
University. Yongqiang Tian is the corresponding author.
\end{acks}

\bibliographystyle{ACM-Reference-Format}
\bibliography{references}

@inproceedings{ko2004whyline,
  author = {Ko, Andrew J. and Myers, Brad A.},
  title = {{Designing the Whyline}: A Debugging Interface for Asking Questions about Program Behavior},
  booktitle = {Proceedings of the SIGCHI Conference on Human Factors in Computing Systems},
  series = {CHI '04},
  year = {2004},
  pages = {151--158},
  publisher = {ACM},
  address = {New York, NY, USA},
  doi = {10.1145/985692.985712},
  url = {https://doi.org/10.1145/985692.985712}
}

@inproceedings{parnin2011debugging,
  author = {Parnin, Chris and Orso, Alessandro},
  title = {Are Automated Debugging Techniques Actually Helping Programmers?},
  booktitle = {Proceedings of the 2011 International Symposium on Software Testing and Analysis},
  series = {ISSTA '11},
  year = {2011},
  pages = {199--209},
  publisher = {ACM},
  address = {New York, NY, USA},
  doi = {10.1145/2001420.2001445},
  url = {https://doi.org/10.1145/2001420.2001445}
}

@inproceedings{sillito2006questions,
  author = {Sillito, Jonathan and Murphy, Gail C. and De Volder, Kris},
  title = {Questions Programmers Ask during Software Evolution Tasks},
  booktitle = {Proceedings of the 14th ACM SIGSOFT International Symposium on Foundations of Software Engineering},
  series = {SIGSOFT FSE '06},
  year = {2006},
  pages = {23--34},
  publisher = {ACM},
  address = {New York, NY, USA},
  doi = {10.1145/1181775.1181779},
  url = {https://doi.org/10.1145/1181775.1181779}
}

@misc{vscodeapi,
  author = {{Microsoft}},
  title = {{Visual Studio Code Extension API}},
  howpublished = {\url{https://code.visualstudio.com/api}},
  year = {2026},
  note = {Accessed 2026-06-24}
}

@book{zeller2009why,
  author = {Zeller, Andreas},
  title = {Why Programs Fail: A Guide to Systematic Debugging},
  edition = {2},
  year = {2009},
  publisher = {Morgan Kaufmann},
  address = {Burlington, MA, USA},
  isbn = {978-0-12-374515-6},
  url = {https://www.sciencedirect.com/book/9780123745156/why-programs-fail}
}

@article{mccauley2008debugging,
  author = {McCauley, Renee and Fitzgerald, Sue and Lewandowski, Gary and
    Murphy, Laurie and Simon, Beth and Thomas, Lynda and Zander, Carol},
  title = {Debugging: A Review of the Literature from an Educational Perspective},
  journal = {Computer Science Education},
  volume = {18},
  number = {2},
  pages = {67--92},
  year = {2008},
  publisher = {Taylor \& Francis},
  doi = {10.1080/08993400802114581},
  url = {https://doi.org/10.1080/08993400802114581}
}

@article{fitzgerald2008flailing,
  author = {Fitzgerald, Sue and Lewandowski, Gary and McCauley, Renee and
    Murphy, Laurie and Simon, Beth and Thomas, Lynda and Zander, Carol},
  title = {Debugging: Finding, Fixing and Flailing, a Multi-Institutional
    Study of Novice Debuggers},
  journal = {Computer Science Education},
  volume = {18},
  number = {2},
  pages = {93--116},
  year = {2008},
  publisher = {Taylor \& Francis},
  doi = {10.1080/08993400802114508},
  url = {https://doi.org/10.1080/08993400802114508}
}

@inproceedings{brown2014blackbox,
  author = {Brown, Neil Christopher Charles and K{\"o}lling, Michael and
    McCall, Davin and Utting, Ian},
  title = {Blackbox: A Large Scale Repository of Novice Programmers' Activity},
  booktitle = {Proceedings of the 45th ACM Technical Symposium on Computer
    Science Education},
  series = {SIGCSE '14},
  year = {2014},
  pages = {223--228},
  publisher = {ACM},
  address = {New York, NY, USA},
  doi = {10.1145/2538862.2538924},
  url = {https://doi.org/10.1145/2538862.2538924}
}

@inproceedings{price2020progsnap2,
  author = {Price, Thomas W. and Hovemeyer, David and Rivers, Kelly and
    Gao, Ge and Bart, Austin Cory and Kazerouni, Ayaan M. and
    Becker, Brett A. and Petersen, Andrew and Gusukuma, Luke and
    Edwards, Stephen H. and Babcock, David},
  title = {ProgSnap2: A Flexible Format for Programming Process Data},
  booktitle = {Proceedings of the 2020 ACM Conference on Innovation and
    Technology in Computer Science Education},
  series = {ITiCSE '20},
  year = {2020},
  pages = {356--362},
  publisher = {ACM},
  address = {New York, NY, USA},
  doi = {10.1145/3341525.3387373},
  url = {https://doi.org/10.1145/3341525.3387373}
}

@inproceedings{spacco2006marmoset,
  author = {Spacco, Jaime and Hovemeyer, David and Pugh, William and
    Emad, Fawzi and Hollingsworth, Jeffrey K. and Padua-Perez, Nelson},
  title = {Experiences with {Marmoset}: Designing and Using an Advanced
    Submission and Testing System for Programming Courses},
  booktitle = {Proceedings of the 11th Annual SIGCSE Conference on Innovation
    and Technology in Computer Science Education},
  series = {ITiCSE '06},
  year = {2006},
  pages = {13--17},
  publisher = {ACM},
  address = {New York, NY, USA},
  doi = {10.1145/1140124.1140131},
  url = {https://doi.org/10.1145/1140124.1140131}
}

@article{wong2016fault,
  author = {Wong, W. Eric and Gao, Ruizhi and Li, Yihao and Abreu, Rui and
    Wotawa, Franz},
  title = {A Survey on Software Fault Localization},
  journal = {IEEE Transactions on Software Engineering},
  volume = {42},
  number = {8},
  pages = {707--740},
  year = {2016},
  publisher = {IEEE},
  doi = {10.1109/TSE.2016.2521368},
  url = {https://doi.org/10.1109/TSE.2016.2521368}
}

@inproceedings{ihantola2015edm,
  author = {Ihantola, Petri and Vihavainen, Arto and Ahadi, Alireza and
    Butler, Matthew and B{\"o}rstler, J{\"u}rgen and Edwards, Stephen H. and
    Isohanni, Essi and Korhonen, Ari and Petersen, Andrew and Rivers, Kelly
    and Rubio, Miguel {\'A}ngel and Sheard, Judy and Skupas, Bronius and
    Spacco, Jaime and Szabo, Claudia and Toll, Daniel},
  title = {Educational Data Mining and Learning Analytics in Programming:
    Literature Review and Case Studies},
  booktitle = {Proceedings of the 2015 ITiCSE on Working Group Reports},
  series = {ITiCSE-WGR '15},
  year = {2015},
  pages = {41--63},
  publisher = {ACM},
  address = {New York, NY, USA},
  doi = {10.1145/2858796.2858798},
  url = {https://doi.org/10.1145/2858796.2858798}
}

@inproceedings{zhou2026bisection,
  author = {Zhou, Xintong and Xu, Zhenyang and Tian, Yongqiang and Sun, Chengnian},
  title = {On the Feasibility of Deduplicating Compiler Bugs with Bisection},
  booktitle = {Proceedings of the 35th ACM SIGSOFT International Symposium on
    Software Testing and Analysis},
  series = {ISSTA '26},
  year = {2026},
  publisher = {ACM},
  address = {New York, NY, USA}
}

\end{document}